\documentclass[twocolumn,preprintnumbers,amsmath,amssymb,groupedaddress,superscriptaddress]{revtex4}
\usepackage{graphicx}
\usepackage{dcolumn}
\usepackage{bm}
\usepackage{amssymb}
\usepackage{amsmath}
\usepackage{epstopdf}
\usepackage{color}
\usepackage{ulem}
\usepackage{soul}
\usepackage[colorlinks=true,citecolor=blue,urlcolor=blue]{hyperref}

\renewcommand{\emph}{\textit}
\begin{document}

\title{Measurement of the excited-state transverse hyperfine coupling in NV centers via dynamic nuclear polarization
}

\author{F. Poggiali}
\affiliation{
LENS European Laboratory for Non linear Spectroscopy, and Dipartimento di Fisica e Astronomia, Universit\`a di Firenze; INO-CNR Istituto Nazionale di Ottica del CNR, I-50019 Sesto Fiorentino, Italy} 
\author{P. Cappellaro}
\affiliation{
LENS European Laboratory for Non linear Spectroscopy, and Dipartimento di Fisica e Astronomia, Universit\`a di Firenze; INO-CNR Istituto Nazionale di Ottica del CNR, I-50019 Sesto Fiorentino, Italy}
\affiliation{Department of Nuclear Science and Engineering, Massachusetts Institute of Technology, Cambridge, MA 02139} 
\author{N. Fabbri}\email{fabbri@lens.unifi.it}
\affiliation{
LENS European Laboratory for Non linear Spectroscopy, and Dipartimento di Fisica e Astronomia, Universit\`a di Firenze; INO-CNR Istituto Nazionale di Ottica del CNR, I-50019 Sesto Fiorentino, Italy}

\begin{abstract}

Precise knowledge of a quantum system's Hamiltonian is a critical pre-requisite for its use in many quantum information technologies.
Here, we report a method for the precise characterization of the non-secular part of the excited-state Hamiltonian of an electronic-nuclear spin system in diamond. The method relies on the investigation of the dynamic nuclear polarization mediated by the electronic spin, which is currently exploited as a primary tool for initializing nuclear qubits and performing enhanced nuclear magnetic resonance. By measuring the temporal evolution of the population of the ground-state hyperfine levels of a nitrogen-vacancy center, we obtain the first direct estimation of the excited-state transverse hyperfine coupling between its electronic and nitrogen nuclear spin. Our method could also be applied to other electron-nuclear spin systems, such as those related to defects in silicon carbide.

\end{abstract}

\maketitle

\section{Introduction}
Negatively charged nitrogen-vacancy centers (NV) in diamond~\cite{DuPreez65} have emerged as promising platforms for quantum information processing~\cite{Ladd10} and for a wide range of applications in quantum sensing~\cite{Degen08, Taylor08, Maze08}.
The NV electronic spin remarkable properties, such as optical initialization and readout of its spin state~\cite{Jelezko06}, and extremely long spin coherence~\cite{Balasubramanian09}, make it an excellent candidate for quantum technologies.
The presence of other nuclear spins in the proximity of the NV defect can be exploited to enhance the quantum computation or sensing tasks, for example to achieve better readout~\cite{Lovchinsky16,Neumann10b}, long-time memory~\cite{Maurer12}, or to implement quantum error correction schemes~\cite{Hirose16,Waldherr14,Cramer16}. A critical step in many of these schemes is to first initialize the nuclear spin in a highly polarized (pure) state~\cite{He93, Fuchs08, Gali09, Smeltzer09, Jacques09}.
 
Polarization of the NV electronic spin to the $m_S = 0$ sublevel of the ground-state spin triplet is routinely obtained via optical pumping and inter-system crossing.
In general, this process does not lead to polarization of the nuclear spin owing to the mismatch between the electron and nuclear spin energies. However, close to the excited state level anticrossing (ESLAC), occurring at magnetic field around 510~G, the transverse hyperfine coupling induces electron-nuclear flip-flops, and consequently polarization transfer from electron to nuclear spins~\cite{Jacques09}. Nearly perfect nuclear polarization has been demonstrated in previous experiments for $^{14}$N~\cite{Dreau12} or $^{15}$N~\cite{Jacques09, Ivady15} composing the NV center, as well as for proximal $^{13}$C~\cite{Smeltzer11, Dreau12, Wang13}.
Recently, dynamic nuclear spin polarization has also been observed in similar defect systems in Silicon Carbide, such as the divacancy in 6H-SiC and the PL6 center in 4H-SiC~\cite{Falk15}.

The polarization transfer dynamics, and its ultimate achievable level, depends critically on the hyperfine spin structure of the ground and excited electronic levels.
Although the spin structures of both the ground~\cite{He93, Rabeau07, Felton08, Fuchs08, Smeltzer09, Doherty13} and excited~\cite{Fuchs08, Neumann09, Steiner10} triplet states have been characterized in experiments, the transverse hyperfine coupling between electronic and nuclear spin is in general difficult to measure. In particular, the excited state transverse hyperfine coupling strength has been inferred by assuming an isotropic interaction~\cite{Fuchs08, Steiner10}, although {\it ab initio} calculations indicate an anisotropy of the hyperfine tensor for the $^{15}$N isotope~\cite{Gali09}.

In this work we design a strategy to measure the excited-state transverse hyperfine coupling, by exploiting dynamic nuclear polarization (DNP) close to the ESLAC. A deeper understanding of this mechanism would allow enhanced control of this multi-spin system, from its initialization to more complex sensing and computational tasks. Our strategy combines measuring the time-dependence of the polarization dynamics with {\it ab initio} calculations based on a master equation in the Lindblad operator formalism~\cite{Lindblad76}.

Comparing the experimental results with the model, we can extract the first experimentally measured value of the transverse hyperfine coupling in the NV electronic excited state.

\section{Polarization mechanism}
We consider the two-spin system given by the electronic spin $S=1$ associated with the NV center, in its orbital ground and excited states, and the nuclear spin $I=1$ of the substitutional $^{14}$N that constitutes the center together with a vacancy in the adjacent lattice site.

\begin{figure}[]
\centering
\includegraphics[width=\columnwidth]{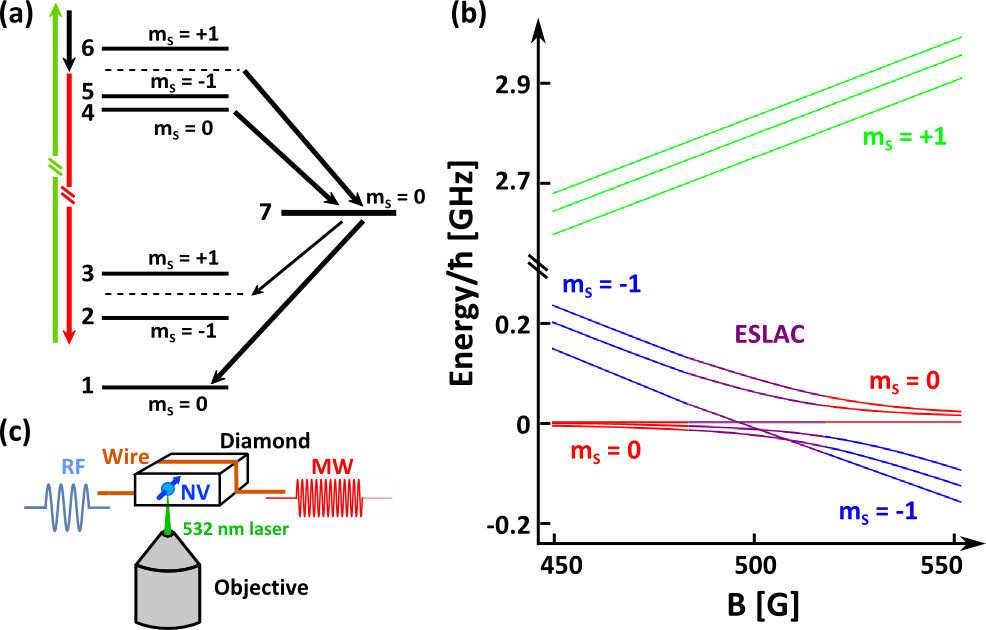}
\caption{\textbf{(a)} Seven-level scheme of the NV electronic structure. Levels 1-3 and 4-6 represent the three different m$_S$ projections of the ground and excited state, respectively. Level 7 represents the electronic singlet metastable level. We show optical excitations at 532~nm (green arrows), radiative decay { within a broad phonon side band in the range 637-800~nm, with zero phonon line at 637~nm (red arrows),} as well as non-radiative decay (black arrows) via the metastable level, responsible for spin polarization. \textbf{(b)} Hyperfine energy levels of the excited state, close to the ESLAC. \textbf{(c)} Sketch of the experimental setup: an objective focuses the excitation laser beam and collects the fluorescence; a wire works as an antenna to deliver MW and RF waves to the NV center and to drive the electronic and nuclear spins, respectively~\cite{SOM}.
}
\label{fig1}
\end{figure}

At room temperature, the orbital ground ($^3A$) and excited ($^3E$) states of the system are governed by the same form of Hamiltonian. Indeed, in the excited state, the orbital contribution to the energy spin levels is quenched due to mixing of the excited state orbital doublet $\{E_x,E_y\}$, attributed to thermally-activated phonon excitations~\cite{Rogers09, Batalov09}. Therefore, the excited state behaves as an effective orbital singlet like the ground state, where spin level energies are determined only by spin-spin and Zeeman interactions. This is no longer the case at cryogenic temperatures, where our model would not apply. A scheme of the level structure generated from these Hamiltonian operators is represented in Fig.~\ref{fig1} (a) and (b).
In the presence of an externally applied magnetic field ${\bf B}$, the excited-state (ES) Hamiltonian reads
\begin{equation}
\label{eq:He}
\mathcal{H}_e = D_e S_z^{2} + \gamma_e~ \mathbf{S} \! \cdot \! \mathbf{B} + Q~ I_z^{2} + \mathbf{S} \! \cdot \! \mathbf{C}\! \cdot \! \mathbf{I} + \gamma_n~ \mathbf{I} \! \cdot \! \mathbf{B}
\end{equation}
where $\mathbf{S}$ and $\mathbf{I}$ are the electronic and nuclear spin operators, $D_e = 1.42$~GHz is the electronic zero-field spitting of the excited state, $Q=-4.945$ MHz is the nuclear quadrupole interaction, $\gamma_e = 2.802$~MHz/G and $\gamma_n = -0.308$~kHz/G are the electronic and nuclear gyromagnetic ratios.
The hyperfine interaction can be rewritten as:
\begin{equation}
\label{eq:Hf}
\mathbf{S} \! \cdot \! \mathbf{C}\! \cdot \! \mathbf{I} = C_{/\!/} S_z I_z + C_{\bot} (S_x I_x + S_y I_y)
\end{equation}
with $C_{/\!/}$ and $C_{\bot}$ the amplitudes of the longitudinal and transverse coupling between the two spins. The ground state Hamiltonian $H_g$ has the same form, with $D_g = 2.87$~GHz and hyperfine coupling tensor $\mathbf{A}$, so that $\mathbf{S} \! \cdot \! \mathbf{A}\! \cdot \! \mathbf{I} = A_{/\!/} S_z I_z + A_{\bot} (S_x I_x + S_y I_y)$. The values of the amplitudes $A_{/\!/}=-2.162$~MHz~\cite{Smeltzer09}, $A_{\bot}=-2.62$~MHz~\cite{Felton08,Chen15} and $C_{/\!/}=-40$~MHz~\cite{Smeltzer09, Steiner10} were experimentally evaluated via electron spin resonance. On the other hand, $C_{\bot}$ has not been experimentally determined and it is often assumed to be equal to $C_{/\!/}$~\cite{Steiner10, Wang13}.

The transverse hyperfine coupling in the excited state is at the basis of the nuclear spin polarization process, since it leads to a mixing of the states with the same total (electronic plus nuclear) spin~\cite{Jacques09}. This mixing becomes relevant near the level anticrossing in the excited state, where $ \left| 0,-1 \right\rangle_e$ mixes with $\left| -1,0 \right\rangle_e$, and $\left| 0,0 \right\rangle_e$ with $\left| -1,1 \right\rangle_e$, as illustrated in Fig.~\ref{fig1} (c). Here, we used the notation $\left| m_S, m_I \right \rangle_e = \left| m_S \right \rangle_e \otimes \left| m_I \right \rangle_e$ to indicate the unperturbed hyperfine levels of the ES, in the absence of couplings and transverse magnetic fields. 
Then, energy-conserving exchange of polarization by spin flip-flop can occur, that, when combined with a continuous cycle of optical excitation and non-radiative decay, leads to a polarization of both the electronic and the nuclear spins. The relative population of the hyperfine levels of the ground-state achieved after long optical pumping depends {\it (i)} on the magnetic field strength and orientation with respect to the NV symmetry axis, and {\it (ii)} on the decay rates of the optical transitions between the spin states (spontaneous emission and intersystem crossing). On the other hand, the temporal dynamics of the nuclear polarization strongly depends on the rate of the flip-flop process, that is, on the transverse hyperfine interaction in the excited state.
Here, we characterize the temporal dynamics of the population of the hyperfine levels in the ground-state of a single NV center, both in experiment and with a theoretical model.
Since the characteristic timescale of the population ({\it resp.}, depletion) of the state $|0,+1\rangle_g$ ({\it resp.}, $|0,0\rangle_g$) crucially depends on the excited-state transverse hyperfine interaction, we can determine the excited-state coupling constant $C_\perp$ with simple magnetic resonance tools.

\section{Experiments}
\label{Experiments}
In the experiment, we used single NVs centers hosted in an electronic grade diamond sample, with natural 1.1\% abundance of $^{13}$C impurities and $^{14}$N concentration $< 5$ ppb (Element Six). The color centers are optically addressed at room temperature with a home-built confocal microscope and their spin was manipulated via resonant microwave driving (Fig.~\ref{fig1} (c)).
The NV centers were chosen to be free from proximal $^{13}$C. We work at magnetic fields ranging from 200~G to 420~G, and with a controlled orientation with respect to the defect symmetry axis. Thus, optical illumination (at wavelength of 532 nm) induces polarization of the nuclear spin with variable efficiency due to the changing proximity to the ESLAC.

At a given magnetic field, we measured the relative population of the hyperfine sublevels of the ground-state electronic spin triplet by performing Ramsey experiments.
We apply two microwave $\pi/2$ pulses, on resonance with the transitions between the spin manifolds ($m_s=0 \leftrightarrow -1$ or $m_s=0 \leftrightarrow +1$), and separated by a variable free evolution time. For each spin transition, three electron spin resonances (ESR) emerge in the Fourier components of the free-evolution signal, corresponding to the three nuclear spin projections $m_I=0,\pm 1$ of $^{14}$N. The typical microwave $\pi/2$ pulse that drives the electronic spin lasts 25-50 ns, with a corresponding Rabi frequency large enough to simultaneously excite all the three transitions separated by the $2.16$~MHz hyperfine interaction~\footnote{The Rabi frequency varies at different driving frequencies, as set by the magnetic field, due to variation in the transmission of the wire used to drive the spins.}.
Due to the high frequency to be probed compared to $1/T_2^*\sim 0.2$~MHz, Ramsey experiments provided high resolution and signal-to-noise ratio.

\begin{figure}[h]
\flushleft
\includegraphics[scale=0.36]{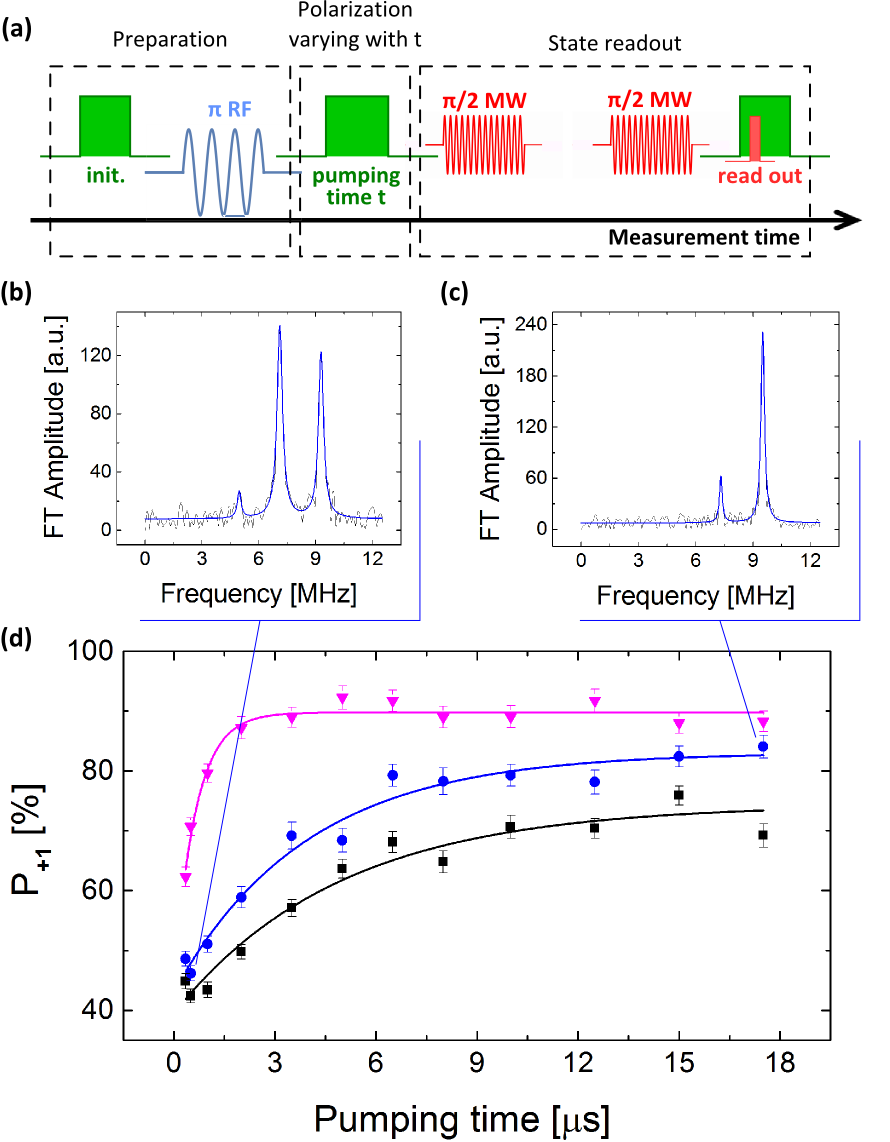}
\caption{\textbf{(a)} Measurement sequence for dynamical nuclear polarization: after an initialization laser pulse, a RF $\pi$ pulse resonant with the $\left|0, +1\right\rangle_g \rightarrow \left|0, 0\right\rangle_g$ transition reverses the two populations; a pumping laser pulse with variable time $t$ re-polarizes the nuclear spin; a Ramsey spectroscopy measurements on the electronic transition $\left|0, m_I\right\rangle_g \rightarrow \left|-1, m_I\right\rangle_g$ evaluates the polarized fraction of the three hyperfine levels. \textbf{(b)} and \textbf{(c)} Fourier transform of the Ramsey measurements for a single NV denoted as NV1 at $B =$ 348~G for pumping time $t = 0.5~\mu$s and $17.5~\mu$s, respectively; blue lines are lorentzian fits. \textbf{(d)} Polarized fraction $P_{+1}$ of the nuclear spin as a function of optical pumping time $t$, obtained from the Ramsey spectra. Black, blue and magenta points corresponds to $(252 G, 1.7^\circ)$, $(348 G, 1.5^\circ)$ and $(411 G, 0.8^\circ)$; the three lines are fit performed with an exponential function $P_{+1} = P_0 - A \, e^{-t/\tau}$.}
\label{fig3}
\end{figure}

Within each spin resonance, the intensities of the different hyperfine transitions give information on the ground state manifold populations (see Fig.~\ref{fig3} (b) and (c)). We extract the relative probability of the nuclear spin projection $m_I$ as:
\begin{equation}
\label{eq:polfracexp}
P_i = \frac{I(\nu_{i})}{\sum_{j} I(\nu_j)}
\end{equation}
where $I(\nu_j)$ is the integral of the Fourier component of the Ramsey signal with frequency $\nu_j$ ($j=0,\pm1$). 

In order to investigate the temporal dynamics of the polarization process, we prepare the system in a mixed state in the lowest-energy electronic level, and then we follow the behavior of polarization under optical illumination of variable time duration at the saturation power (see Fig.~\ref{fig3} (a)). For the preparation, first a 20 $\mu$s-long optical excitation partially polarizes the NV-$^{14}$N system, driving it into an unbalanced mixed state $\alpha_{-1} |0,-1\rangle\!\langle0,-1|_g + \alpha_0 |0,0\rangle\!\langle0,0|_g + \alpha_{1} |0,1\rangle\!\langle0,1|_g$, where $\alpha_{1} \sim 1$ for fields close to the ESLAC, and $\alpha_{i}$ depend on the magnitude $B$ of the external magnetic field and on the angle $\theta$ with the NV axis. Then, a radiofrequency $\pi$ pulse ($t_\pi\sim 30$ $\mu$s) on resonance with the $|0,+1\rangle_g\leftrightarrow|0,0\rangle_g$ coherently reverses the population of nuclear spin projections $m_I=0,+1$ and alters polarization.
To reveal the polarization dynamics, we use an optical pulse of variable length $t$, and probe the resulting population of the hyperfine levels with the Ramsey experiment explained above.
We characterize the polarization dynamics for different values of the magnitude and different orientations of the magnetic field. The method used for calibrating magnitude and orientation of the the magnetic field is described in Appendix~\ref{calibration}. The polarized fraction $P_{+1}$ is reported in Fig.~\ref{fig3} (d) as a function of the optical pumping time $t$ for $(B,\theta)= (252$~G, $1.7^\circ)$, $(348$~G, $1.5^\circ)$ and $(411$~G, $0.8^\circ)$. We observe that $P_{+1}$ increases in time until reaching its final value, with variable time-constant ranging from 1 to 5~$\mu$s. This saturation level corresponds to the equilibrium condition between the two competing processes: flip-flop between electronic and nuclear spin and optical spin pumping.

\section{Numerical model}
\label{Numerical}
We compare the experimental results with simulations obtained by modelling the time evolution of the two-spin state with the Master equations in the Lindblad form~\cite{Fischer13b,Ticozzi09}. In turns, this allow us to determined the unknown parameters in the model.

The time evolution is dictated by the ground-state and excited-state Hamiltonians ($H_g$ and $H_e$, which generate a coherent dynamics) as well as Markovian processes associated with coupling to photons and phonons, that induce transitions between different spin and orbit configurations, such as laser excitation, spontaneous and stimulated emissions, as well as intersystem crossing.

The two-spin system is described by the density operator $\rho$ consisting of 21 hyperfine states -- 9 in the ground state, 9 in the excited state, and 3 in the singlet state. We calculate the population of the hyperfine sublevels of the ground state and the polarized fraction from the diagonal elements of the density matrix.

The time evolution of $\rho$ is described by the generalized Liouville equation:
\begin{equation}
\label{eq:liouville}
\frac{d}{dt}\rho = - \frac{i}{\hbar}[\mathbf{H},\rho] + \hat{L}[\rho]
\end{equation}
with $\mathbf{H}$ the total spin Hamiltonian of ground and excited states.
This master equation allows us to go beyond a simple rate equation model, and fully account for the effects of transverse fields as well as coherent spin polarization exchange.
The Lindbald operator $\hat{L}$ in the second term on the right is related to jumps $L_k$ between different spin states through the equation~\cite{Lindblad76}:
\begin{equation}
\label{eq:liouville2}
\hat{L}[\rho] = \sum_{k=1}^{N} \left( L_k \rho(t)L_k^\dagger - \frac{1}{2}L_k^\dagger L_k \rho(t) - \frac{1}{2} \rho L_k^\dagger L_k \right)
\end{equation}
Most generally, we can write the jump operators as $L_k = \sqrt{\Gamma_{mn}} \left| m \right\rangle\!\left\langle n \right|$, with $\Gamma_{mn}$ the rate of the transition between $\left| m \right\rangle$ and $\left| n \right\rangle$.
We consider spin-conserving radiative transitions and the decay from the excited states to the ground through the metastable $S = 0$ level. We also introduce the contribution of spin non-conserving radiative processes, the rate of which we evaluated as $\epsilon=0.01$ of the rate of spin conserving transitions~\cite{Robledo11b}. All the rates related to these transitions are reported in Table~\ref{tab:gamma}. Note that these parameters have been independently measured before, from the dynamics of the NV center electronic spin alone~\cite{Manson06,Robledo11b,Tetienne12}. In order to reproduce the measured polarization evolution at saturation, and extract the strength of the transverse hyperfine coupling from the comparison between theory and data, we set the optical pumping rate equal to the corresponding radiative relaxation rate.
In our model, we neglect the NV ionization process during optical illumination, which we demonstrate to give a small correction of the calculation, as discussed in Appendix \ref{csc}.
\begin{table}[h!]
\caption{Transitions and decay rates (from~\cite{Robledo11b}). The labels correspond to the energy levels in Fig.~\ref{fig1}(a).}
\begin{center}
\begin{tabular}{l c c}
\hline
\hline
Transition & & Rate [MHz]\\ 
\hline
\hline
\\
\vspace{0.2cm}
Spontaneous Emission & $\Gamma_{41}$, $\Gamma_{52}$, $\Gamma_{63}$ & $63 \pm 3$\\
Intersystem crossing & $\Gamma_{47}$ & $12 \pm 3$\\
\vspace{0.2cm}
from ES to singlet& $\Gamma_{57}$, $\Gamma_{67}$ & $80 \pm 6$\\
Intersystem crossing& $\Gamma_{71}$ & $3.3 \pm 0.4$\\
\vspace{0.2cm}
from singlet to GS& $\Gamma_{72}$, $\Gamma_{73}$& $2.4 \pm 0.4$\\
\hline
\hline
\end{tabular}
\par
\end{center}
\label{tab:gamma}
\end{table}

The only experimentally unknown parameter in our model is then the transverse coupling $C_{\bot}$, that influences the rate of the flip-flop process and therefore determines the DNP dynamics.

\begin{figure}[b]
\centering
\includegraphics[width= .9 \columnwidth]{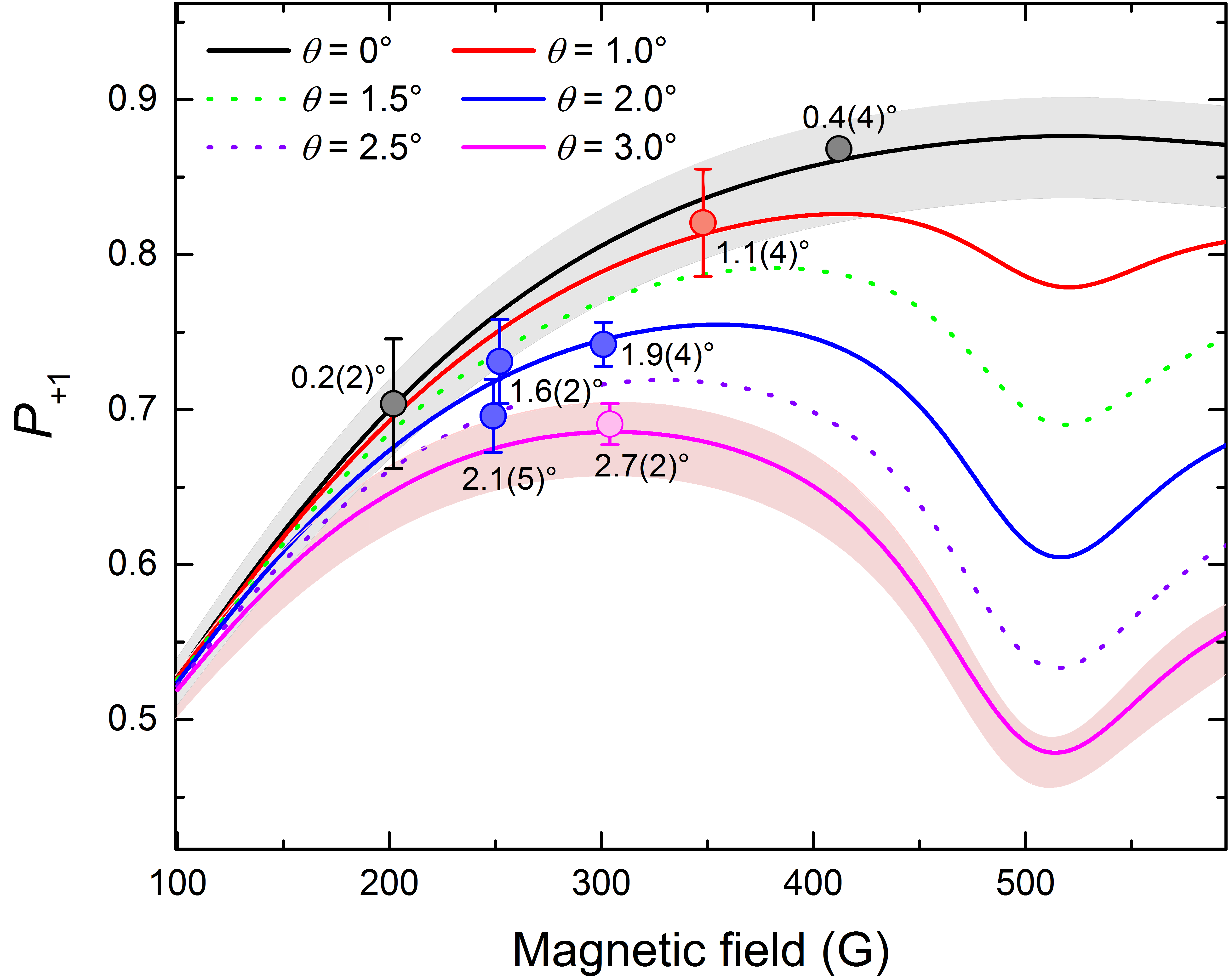}
\caption{Steady-state population $P_{+1}$ of the hyperfine state $|0,+1\rangle_g$, as a function of the modulus of the magnetic field, $B$, for different angles $\theta$ with respect to the NV axis. The curves are numerical solutions of the generalized Liouville equation ($\theta= 0.0^\circ, 1.0^\circ, 1.5^\circ, 2.0^\circ, 2.5^\circ$, and $3.0^\circ$). For the $\theta=0^\circ,3^\circ$ lines, the shaded area represents the error due to the uncertainty in the decay rates reported in Table~\ref{tab:gamma} (we expect similar uncertainties for the other angles). Circles are experimental results, with color code and labels indicating the field orientation $\theta$.}
\label{fig-steady-vs-B}
\end{figure}

With these mathematical tools, we performed numerical simulations in different temporal regimes of the optical pumping, investigating both the transient behavior for short time durations, and the stationary case. We first find, both experimentally and in simulations, that long optical pumping leads to a maximum constant polarized fraction, which depends on the magnetic field amplitude and its orientation with respect to the NV axis.
Comparing the asymptotic polarization obtained from simulation and from experiments allowed us to verify the validity of our model. We note that our model reproduces very well the experimental findings at small angles ($\theta < 3^\circ$), as shown in Fig.~\ref{fig-steady-vs-B}. For larger angles the observed polarization is lower than expected; this deviation could be attributed to other spin decoherence processes in the excited state that reduce the effective interaction time available for the polarization exchange~\cite{Gali09}. Although in our model we did not include these processes, such as the excited state electronic spin dephasing, we verified that they do not have a significant influence on the dynamics at small angles.

Once defined the model that can reproduce well the behavior of the nuclear spin polarization for long polarization times, we investigate the dynamics of the process and its characteristic times.

\begin{figure*}[]
\centering\includegraphics[width= .8 \textwidth]{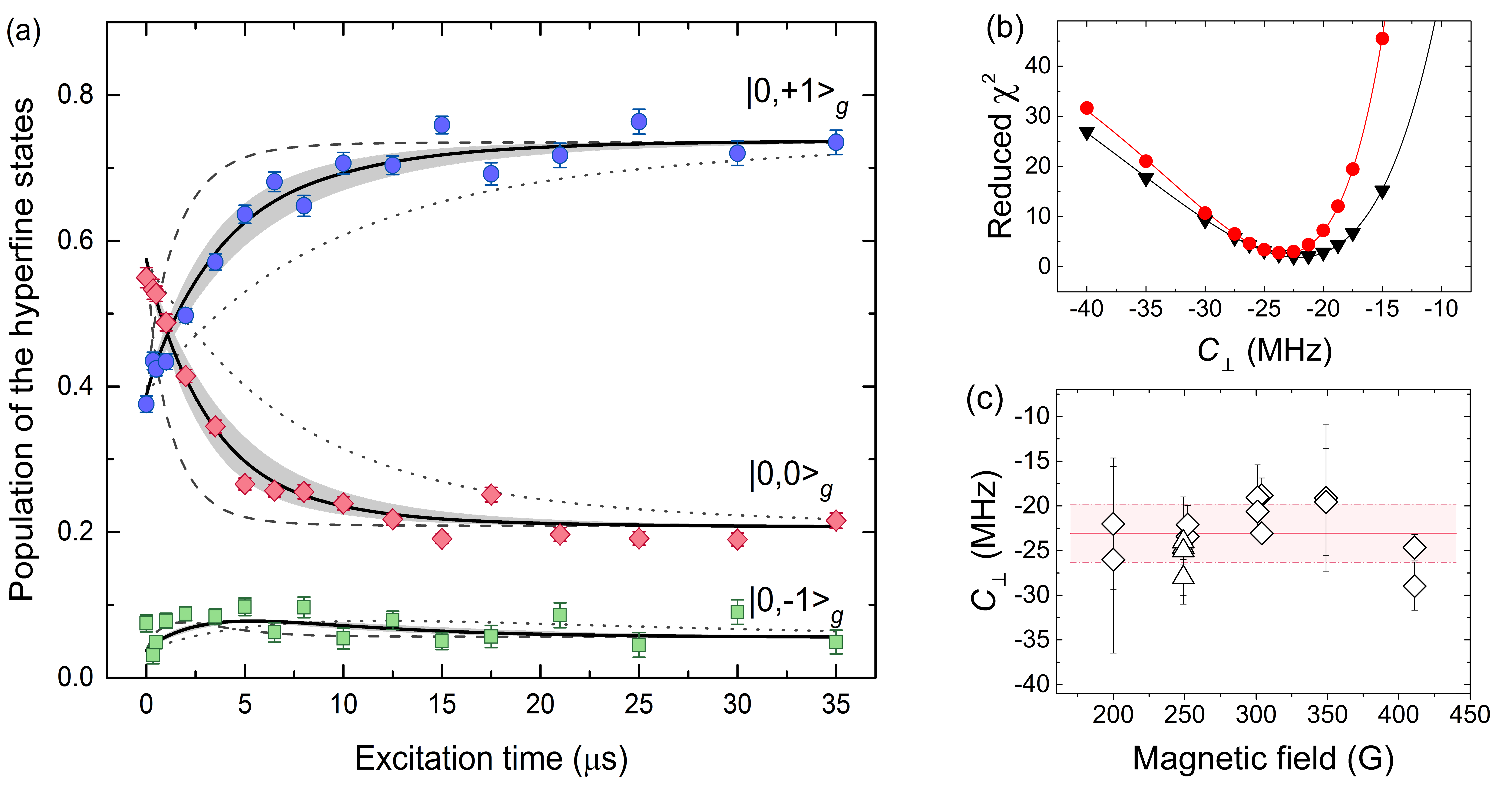}
\caption{\textbf{(a)} Comparison between experimental data and calculation of the relative population of the states $|0,-1\rangle_g$, $|0,0\rangle_g$ and $|0,+1\rangle_g$ after optical pumping of variable length $t$. Blue dots, red diamonds, and green squares correspond to $m_I = +1,0,-1$ nuclear spin relative probability. Dotted and dashed lines are the theoretical curves for $C_{\bot} = -15$~MHz and $-40$~MHz, respectively; black line and gray region correspond to $C_{\bot} = (-23 \pm 3)$~MHz.
\textbf{(b)} Mean squared residuals $\chi^2$ between data and theoretical curves, as a function of the hyperfine transverse coupling $C_{\bot}$ at $B=252$ G; black and red scatters refer to $m_I = +1$ and $m_I = 0$, respectively. The fit to the residuals (black and red lines) were used to find the minimum of the residual distribution and thus the best-fit estimate for $c_\perp$.
\textbf{(c)} Transverse hyperfine coupling parameter of the excited state, $C_\perp$, evaluated for different values of the magnetic field. The analysis of both the $|0,+1\rangle_g$ and the $|0,0\rangle_g$ components is included, for NV1 (diamonds) and NV2 (triangles).
Red straight line and shaded region denote weighted average and standard deviation of the sixteen values of ($B, \theta$).}
\label{Fig-Cperp}
\end{figure*}

\section{Discussion}
\label{Discussion}

We now discuss the time-evolution the population of the $|0,+1\rangle_g$ and $|0,0\rangle_g$ states as a function of the interaction between the optical excitation and the NV system.

The relative population of the nuclear spin projection $P_i$ at long-polarization time strongly depends on the angle between the magnetic field and the symmetry axis.
We note that the other independently evaluated parameter, the magnetic field modulus $B$, affects less crucially the polarization level for uncertainties of the order of few Gauss, which is our case. Similarly, the parameter we want to estimate, $C_\perp$, does not determine the asymptotic polarization (which can then be used to estimate the magnetic field angle, as explained in Appendix~\ref{calibration}), but it affects dramatically the timescale of the polarization dynamics.

For each experimental condition, $B$ and $\theta$, we performed simulations of the time-evolution of the state probability as a function of $C_{\bot}$, which is the only free parameter in the master equation.
This was done for both the $|0,+1\rangle_g$ and the $|0,0\rangle_g$ spin components. The $|0, -1\rangle_g$ was excluded because in most cases the amplitude of its Ramsey component is very small and comparable with our signal to noise ratio.
In Fig.~\ref{Fig-Cperp}~(a) we report the relative probability of the states $m_I= 0,\pm 1$ as a function of the optical pumping time for $B = 252$~G, compared with the theoretical calculation for $C_{\bot} = - 15$~MHz, $- 23$~MHz, $- 40$~MHz. We note that the value often used in literature, $C_{\bot}=-40$~MHz~\cite{Steiner10, Wang13}, which derives from the assumption of isotropic interaction in the excited state, does not fit the experimental findings -- neither the rise-time of the population of the $|0,+1\rangle_g$, or the-decay time of the $|0,0\rangle_g$ population.

For both $|0,+1\rangle_g$ and $|0,0\rangle_g$, we analyze the mean squared residuals, $\chi^2$, between data and theoretical curves, as shown in Fig.~\ref{Fig-Cperp}~(b): the residuals were then fitted with an empirical function~\cite{SOM} to evaluate the best-fitting $C_{\bot}$. By averaging over the two nuclear spin components and over the different experimental magnetic field magnitudes and orientations (Fig.~\ref{Fig-Cperp}~(c)), we obtain a precise estimate of the transverse hyperfine coupling, $C_{\bot} = (-23 \pm 3)$~MHz. This is the first experimental measurement of the transverse hyperfine coupling in the excited-state of the NV center. The common assumption $C_{\bot}$, stemming from the measurement of the secular coupling constant $C_{/\!/}$, is not consistent with the present experimental observation of the timescale of the nuclear polarization.

\section{Concluding remarks}

In conclusion, we have explored the temporal dynamics of nuclear spin polarization of an electron-nuclear hybrid spin system, composed by a single NV center and its $^{14}$N nuclear spin. We found that the timescale of the polarization in the sublevel $|0,+1\rangle_g$ of the ground-state hyperfine triplet (and simultaneous depletion of the $|0,0\rangle_g$ state) crucially depends on the excited-state transverse hyperfine interaction. Exploiting this dependence, we have reported the first precise experimental estimation of the excited-state hyperfine coupling constant $C_\perp$ with simple magnetic resonance tools, obtaining a better knowledge of the nonsecular parts of the system Hamiltonian in the excited state. Our result does not depend on the specific NV, and is representative of NVs in low concentration bulk diamond. Our findings can be useful in NMR experiments enhanced by DNP, hyperpolarization of nuclear spin ensembles, and in all the protocols involving fast and accurate control of nuclear spins, which are crucial for many applications in quantum technologies, including quantum computation, communication and sensing.

\acknowledgements
This work was supported by EU-FP7 ERC Starting Q-SEnS2 (Grant n. 337135), and LaserLab-Europe (Grant Agreement n. 284464, EC's SeventhFramework Programme).

\appendix
\section{Magnetic Field Calibration}
\label{calibration}
\begin{figure}[t]
\centering\includegraphics[width= .9 \columnwidth]{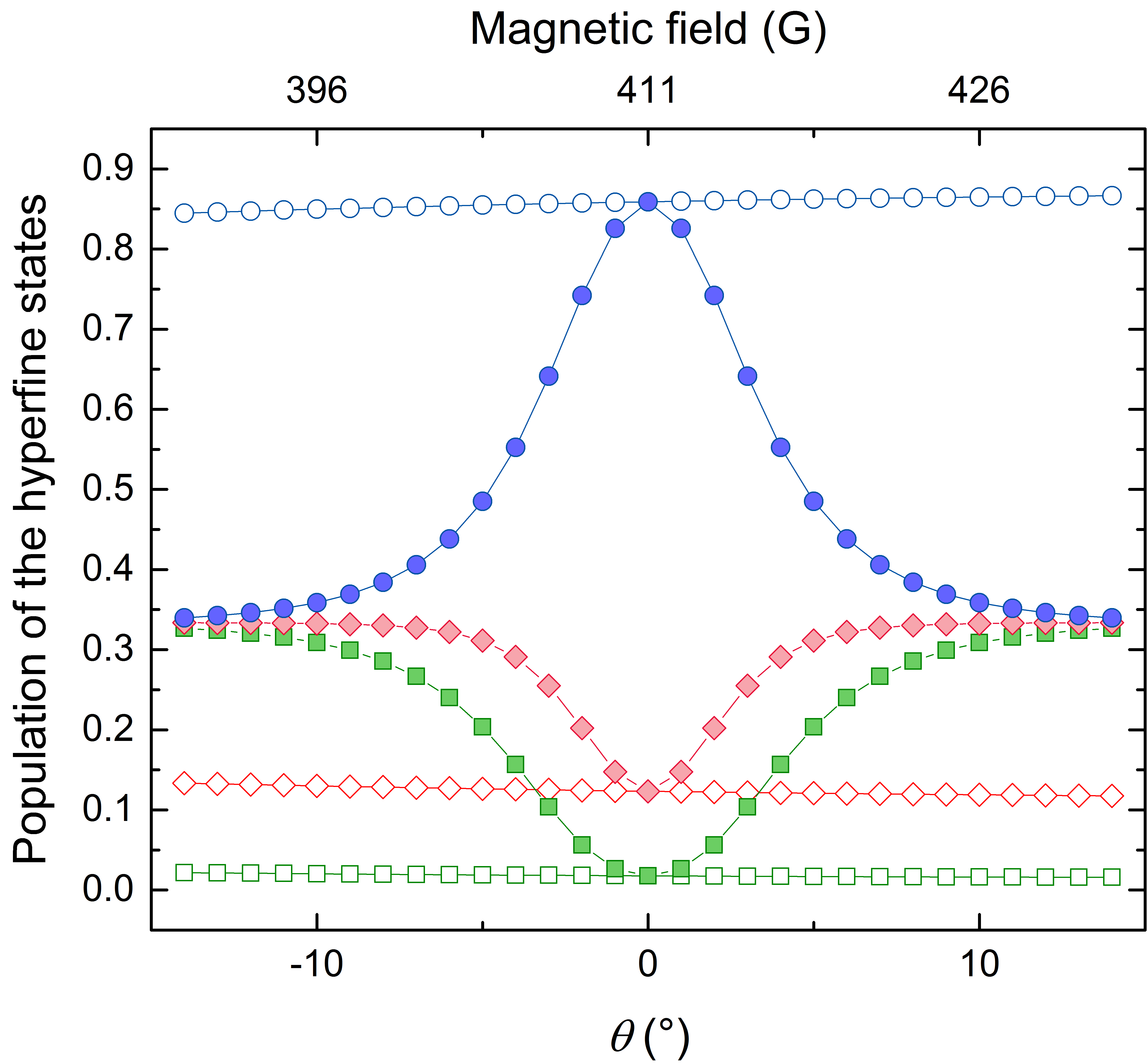}
\caption{Steady-state population of the nuclear spin projections $m_I=0,\pm 1$ in the presence of a magnetic field as a function of the orientation angle $\theta$ with fixed magnitude $B = 411$~G (solid symbols: $m_I=+1$ blue dots, $m_I=0$ red diamonds, $m_I=-1$ green squares), and as a function of the magnitude $B$ in the range $B =390-432$~G for $\theta = 0°$ (empty symbols, same color code).
}
\label{fig-steady}
\end{figure}
Here, we detail the methods used to calibrate the magnitude $B$ of the local magnetic field at the NV position, and the angle $\theta$ between the magnetic field and the symmetry axis of the system.

The first calibration method relies on the measurement of the resonance frequencies $\nu_{\pm}$ of the two ground-state spin transitions $m_S = 0 \rightarrow \pm1$, which are univocally determined by $B$ and $\theta$.

In the presence of the magnetic field $\mathbf{B}$, and neglecting the hyperfine coupling, the ground-state Hamiltonian of the NV electronic spin $S=1$ can be written as~\cite{Doherty13}:
\begin{equation}
\label{eq:Hg}
\mathcal{H}_g = \gamma_e~ \mathbf{S} \! \cdot \! \mathbf{B} + D_g (\hat{S}_z^{2} - 2/3) + E(\hat{S}_x^2 - \hat{S}_y)^2,
\end{equation}
where $\hat{S}_x$, $\hat{S}_y$, $\hat{S}_z$ are the spin operators, $D_g = 2.87$~GHz is the fine structure splitting, and the parameter $E$ is related to strain~\cite{Fuchs08, Neumann09}. For NV centers in ultrapure bulk diamond with low nitrogen concentration, as those investigated in the present work, $E \ll D_g$ and it can be neglected. Thus, the eigenvalues of the Hamiltonian can be found as the solutions of the following characteristic equation
\begin{equation}
\label{eq:eigenHg}
\lambda^3 - 2 D_g \lambda^2 + (D_g^2 - (\gamma_e B)^2) \lambda + \frac{D_g}{2} (\gamma_e B)^2 \left(1 - \cos(2 \theta) \right) = 0
\end{equation}
and depend on $B$ and $\theta$. Note that we work at local magnetic fields well-aligned with the NV symmetry axis, and far away from ground-state level-anticrossing (GS-LAC) occurring at around $B\sim1025$~G, so that the eigenvalues correspond to well defined electronic spin projections $m_S = 0,\pm 1$. We directly evaluated the zero-field splitting $D_g$ with a magnetic resonance experiment in the absence of any external static magnetic field, and measure the two frequencies $\nu_\pm(B,\theta)$ of the transitions $m_S = 0 \rightarrow \pm1$ of the NV in the local magnetic field via Ramsey spectroscopy. Then, we obtain a set of two equations $\nu_\pm(B,\theta) = (E_{\pm 1}(B,\theta) -E_0(B,\theta))/h$, where $E_{0,\pm 1}$ are the eigenvalues of the ground-state Hamiltonian, determined by Eq.~\ref{eq:eigenHg}. These equations can be solved with respect to the two unknown parameters $B$ and $\theta$. Because of uncertainty in the measured $D_g$ value, as well as the measured frequencies, the estimate of the magnetic field angle $\theta$ is not accurate enough.
\begin{figure}
\centering
\includegraphics[width= \columnwidth]{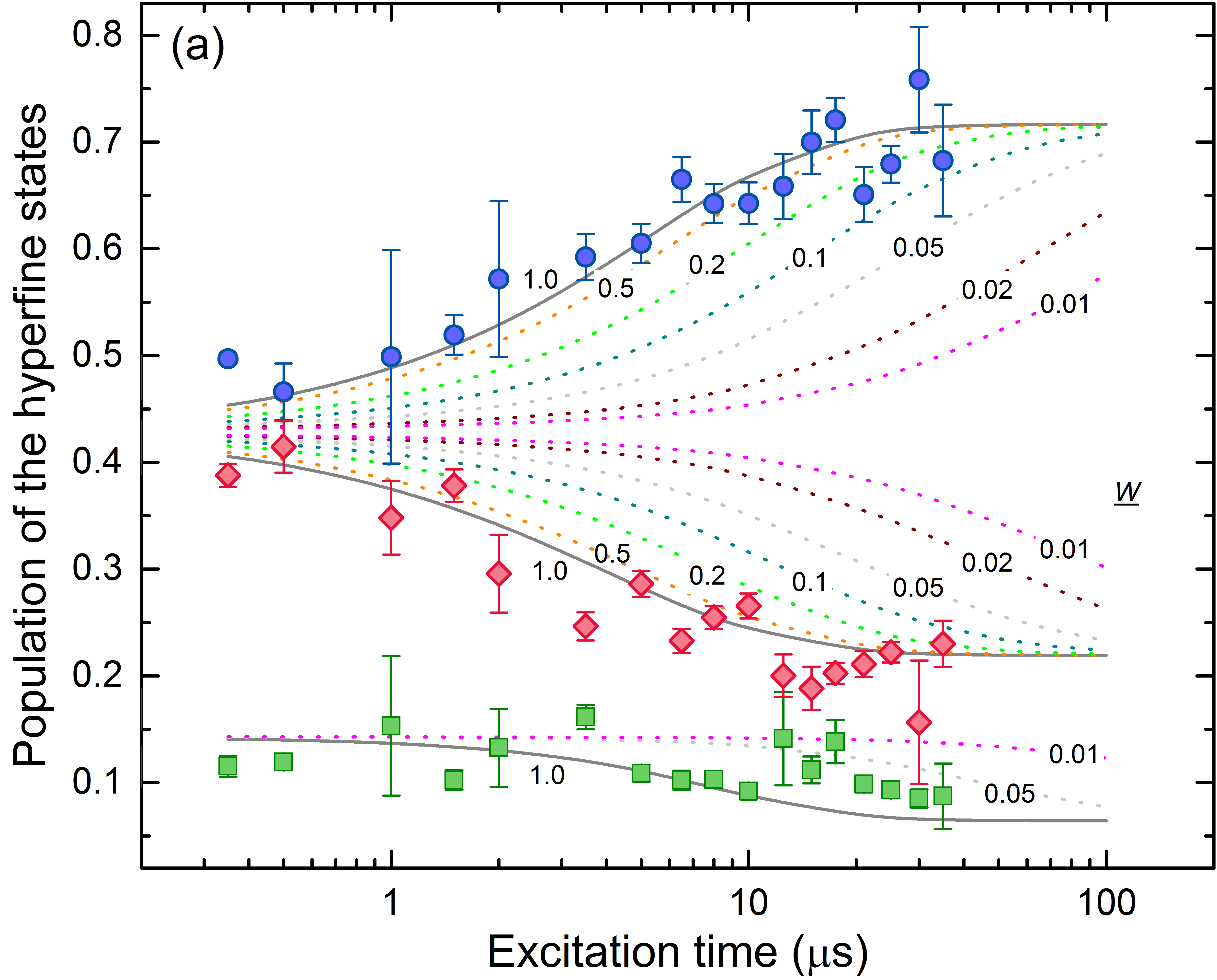}
\includegraphics[width= .7 \columnwidth]{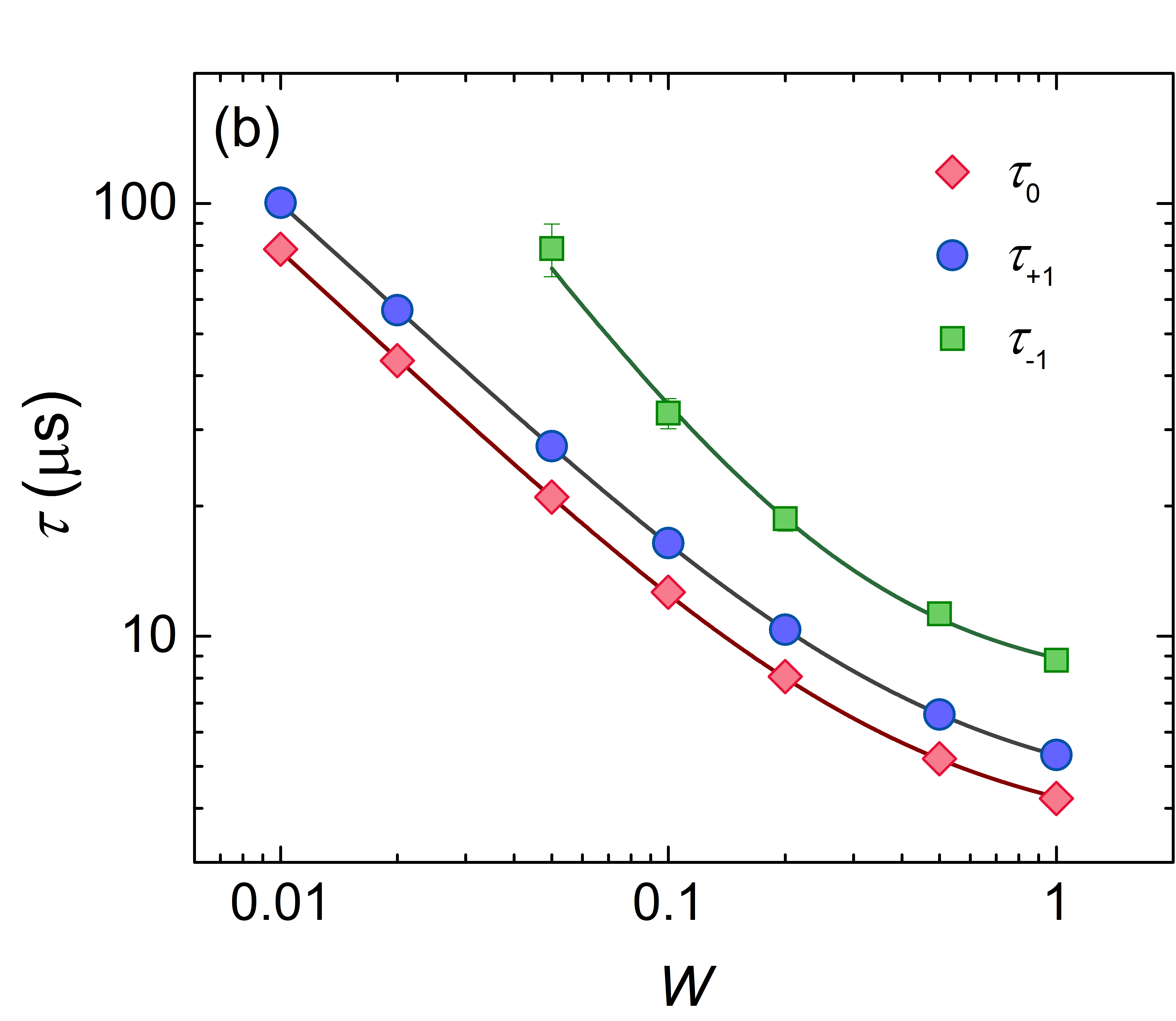}
\caption{(a) Time-evolution of the populations of the hyperfine sublevels of the electronic ground-state, for a magnetic field $B=249$~G aligned along $\theta= 2.1^\circ$. The points are the experimental results obtained for NV2, with laser excitation at saturation power (blue dots, population of $|0,+1\rangle_g$; red diamond, $|0,0\rangle_g$; green squares, $|0,+1\rangle_g$).
The curves are obtained from the solutions of the generalized Liouville equation with $C_{\perp}=-23$~MHz, for different excitation rates $W\Gamma$. The solid line corresponds to $W = 1$, {\it i.e.}, to the excitation rate at saturation power. The dotted lines refer to $W = 0.01, 0.02, 0.05, 0.1, 0.2$, and $0.5$, as denoted in the Figure. For $W\geq1$, we do not observe further changes in the population dynamics.
(b) Characteristic rise-time $\tau_{+1}$ of the population of the state $|0,+1\rangle_g$ (blue dots), and decay-times $\tau_{0}$ and $\tau_{-1}$ of the population of the state $|0,0\rangle_g$ (red diamonds) and $|0,-1\rangle_g$ (green squares) as a function of the pumping rate parameter $W$, in log-log scale. The curves are logarithmic fit of $\tau_{0,\pm1}$.}
\label{fig-power}
\end{figure}

Thus, we also extract an independent estimate of the orientation angle $\theta$ of the local magnetic field, by measuring the steady-state populations $P_{0,+1}$ of the nuclear spin projections $m_I = 0,+1$ of the spin state $m_S=0$. The solution of the generalized Liouville equation has shown us that the short-time dynamics of the hyperfine populations is governed by the excited-state transverse hyperfine coupling $C_\perp$, as discussed in Sec.~\ref{Discussion}, whereas the steady-state populations $P_{0,+1}^\infty$ at long-polarization time is unaffected by $C_\perp$, as shown in Fig.~\ref{Fig-Cperp}~(a). The steady-state populations $P_{0,+1}^\infty$ is found instead to strongly depend on the angle $\theta$, and less crucially on the magnitude of the field $B$, within our typical experimental uncertainty of the order of few \%, as exemplified in Fig.~\ref{fig-steady}.

For any given $B$ and $\theta$, we evaluate the characteristic rise-time $\tau$ ({\it resp.}, depletion time) of $P_{1}(t)$ ({\it resp.}, $P_{0}(t)$). The time average of $P_{0,+1}(t)$ for $t>5\tau$ is used to estimate $P_{0,+1}^\infty$. The theoretical steady-state populations $P_{0,+1}^\infty(B, \theta)$ is fitted to the experimental data, with the angle $\theta$ as the only free parameter of the fit, by minimizing the mean squared residuals $\chi^2$ between data and theoretical curves.

With this second method, we extracted a refined estimate of the angle $\theta$, which we found to be consistent with (but more accurate than) the value estimated from the frequencies of ground-state spin transitions.
We use this refined estimate of the angle as an input in further calculations of the polarization dynamics.

\section{Effects of the Laser Excitation Power}
\label{power}

The excitation rate from the ground to the excited levels, set by the optical power, strongly influences the time evolution of the population of the hyperfine sublevels of the electronic ground-state. To discuss its role, we introduce the optical pumping parameter $W$. Since we consider the relaxation rate via the spin-conserving radiative decay channel to be spin-independent (i.e., $\Gamma=\Gamma_{41}=\Gamma_{52}=\Gamma_{63}$), and the optical pumping rates from the ground to the excited level to be proportional to the corresponding relaxation rates~\cite{Tetienne12}, we define $W=\Gamma_{ij}/\Gamma_{ji}$, with $i=1,2,3$, and $j=4,5,6$. 
In Fig.~\ref{fig-power} we characterize the time evolution of the populations $P_{0,+1}$ of three hyperfine states $|0,0\rangle_g$ and $|0,\pm1\rangle_g$ as a function of $W$, for a typical value of magnetic field ($B=249$ aligned along $\theta= 2.1^\circ$).
Figure~\ref{fig-power}~(a) shows the theoretical curves obtained from the solutions of the generalized Liouville equation described in Sec.~\ref{Numerical} with $C_{\perp}=-23$~MHz, and compare them with the experimental data obtained with optical excitation at the saturation power. From the exponential fit of the theoretical curves we obtain the characteristic rise-time $\tau_{+1}$ of the population of the hyperfine state $|0,+1\rangle_g$, and the decay-times $\tau_0$ and $\tau_{-1}$ of the population of $|0,0\rangle_g$ and $|0,-1\rangle_g$, respectively. We observe that the characteristic times $\tau_{0,\pm1}$ drop logarithmically when increasing the pumping parameter $W$, as shown in Fig.~\ref{fig-power}~(b).

We stress that all the experiments discussed in Sec.~\ref{Discussion}, and used to extract the strength of the transverse hyperfine coupling, were performed by exciting the NV defect at the saturation power, and simulations were conducted setting the optical pumping rate to be equal to the corresponding relaxation rate, in order to reproduce the measured time evolution of the hyperfine sublevels.\\

\section{Effects of charge-state conversion dynamics}
\label{csc}
\begin{figure}[htb]
\centering
\includegraphics[width=0.48\textwidth]{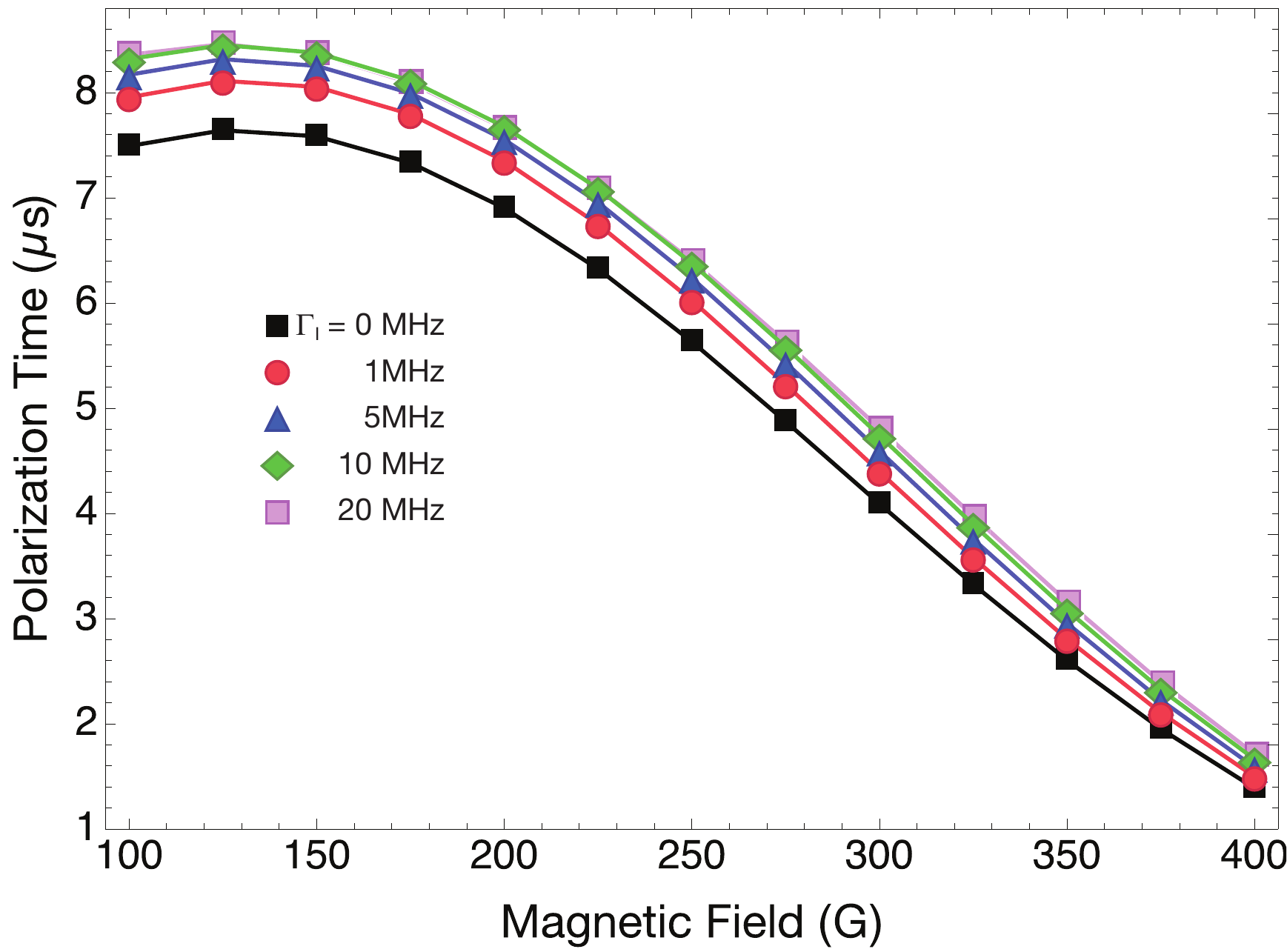}
\caption{Characteristic rise-time $\tau_{+1}$ of the nuclear polarization process as a function of the magnetic field strength for different ionization-recombination rates, $\Gamma_I =0$ (black squares), 1 (red circles), 5 (blue triangles), 10 (green diamonds), and $20$~MHz (pink, light squares).}
\label{fig-ionization}
\end{figure}
The negatively charged NV center can undergo ionization (charge-state conversion to NV$^0$) during the 532~nm laser excitation. The ionization-deionization process has been studied under various conditions of laser wavelength and power~\cite{Grotz12, Shields15,Hopper16}, with rates, related to the excitation powers, varying between kHz~\cite{Aslam13,Chen13b} and MHz~\cite{Chen15p}.
The NV$^-$-NV$^0$ transitions during the polarization laser pulse can reduce the efficiency of the nuclear polarization mechanism and slow down its dynamics. To investigate the contribution of these effects on our estimate of $C_\perp$, we added in our simulations a simple model of the transitions involving the NV$^0$ state. The ionization process can only occur from the NV$^-$ excited states, and the NV$^0$ state then decays to the NV$^-$ ground state~\cite{Aslam13}; we assumed that these transitions are nuclear-spin conserving~\cite{Dhomkar16} and have nuclear-spin-independent rates. We characterize the nuclear polarization time $\tau_{+1}$ as a function of the magnetic field strength for different ionization-recombination rates $\Gamma_I$. The results of the simulations are shown in Fig.~\ref{fig-ionization}. We observe that $\tau_{+1}$ increases with $\Gamma_I$, more markedly at relatively low magnetic field ($B\sim 100-200$~G) than close to the ESLAC, and saturates for $\Gamma_I \gtrsim 10$~MHz. We also evaluated $C_\perp$ including the ionization process for $B = 252$ and $348$~G, and by fitting with the experimental data we find in both cases a small decrease of $C_\perp$. The decrease, as expected, is higher (up to $5$~MHz) for $\Gamma_I = 10$~MHz. Even for this large ionization rate, the estimated $C_\perp$ values are compatible with their values in the absence of ionization, given our estimate uncertainty. At saturation power, we can assume $\Gamma_I$ to be on the order of 1~MHz, implying a correction of $\sim 5\%$ of our estimate of the transverse hyperfine coupling, much smaller than our experimental uncertainty.
Finally, we note that if the ionization process had a larger effect, we would see a more pronounced effects at lower fields, where the polarization times were much longer than at higher field. This would have lead to a variation of the estimated $C_\perp$ with the magnetic field strength, which is instead absent, as shown in Fig.~\ref{Fig-Cperp}(c).

\bibliographystyle{apsrev4-1}
\bibliography{Biblio}
\end{document}